\documentclass[12pt, a4paper]{article}

\usepackage{graphicx, bbold,mathtools,siunitx, float} 
\usepackage{tensor}
\usepackage{dirtytalk}

\usepackage[
    backend=biber,
    style=numeric,    
    sorting=none      
]{biblatex}
\usepackage{hyperref}
\usepackage{authblk} 
\usepackage{mathrsfs}
\title{Microcanonical phase space and entropy in curved spacetime}

\author{Avinandan Mondal\thanks{Current Affiliation: \textit{Raman Research Institute, Sadashivanagar, Bengaluru 560080}. Email: \texttt{avinandan@alumni.iitm.ac.in}},\hspace{2mm} Dawood Kothawala\thanks{Email: \texttt{dawood@iitm.ac.in}}}
\affil{\textit{Department of Physics, Indian Institute of Technology Madras, Chennai 600036.}}

\date{} 

\begin{document}

\maketitle
\begin{abstract}
\noindent We discuss the structure of microcanonical ensembles in inertial and non-inertial frames attached to a confined system of positive energy particles in curved spacetime. Under certain physically reasonable assumptions that ensure the existence of such ensembles, we obtain, for microcanonical ensembles, exact analytical results in certain stationary spacetimes such as Rindler, Schwarzschild, and de Sitter along with leading curvature corrections in arbitrary curved spacetimes. For de Sitter, the exact results have interesting limits when the size of the system is comparable to ${\Lambda}^{-1/2}$. We further highlight two generic characteristics of the leading curvature corrections for a point particle system confined to a spherical or cubical box: (1) they are characterized by Ricci and Einstein tensors, and (2) their contribution is proportional to the bounding area. We argue that the area scaling in (2) does not hold for arbitrary box geometries. We also present a general argument to highlight two distinct sources of divergences in the phase space volume, coming from redshift and spatial geometry, and illustrate this by comparing and contrasting the results for (i) geodesic box in de Sitter, (ii) geodesic box in Schwarzschild, and (iii) uniformly accelerated box in Minkowski. Finally, we extend these results to $N$ particle systems in the restricted case of massless (ultra-relativistic) limit for static spacetimes, for which the results follow very simply from single particle results. Furthermore, we show that the ultra-relativistic expression for equipartition of energy in flat spacetimes continues to hold in static spacetimes.

\vspace{1em}
\end{abstract}

\tableofcontents

\section{Introduction}
Statistical mechanics of particle systems in relativistic settings and self-gravitating systems has been a long-standing topic of interest from both a purely academic perspective and for its relevance in astrophysical situations. Extensive discussions on this topic exist in the literature, and some notable contributions include works on isothermal spheres in Newtonian gravity \cite{PaddyIso, Lynden}, relativistic kinetic theory \cite{Alba}, and self-gravitating quantum gases \cite{Pierre}, among others. The unique issues posed by gravity due to its long range nature and, in the general relativistic framework, due to self-gravitating particles back reacting on the metric, are well acknowledged in the community, and it is expected that a satisfactory clarification of some of these issues might be of fundamental significance. For instance, negative specific heat capacities associated with self-gravitating systems like isothermal spheres \cite{Lynden} and black holes \cite{Hawking} highlight the inequivalence of statistical ensembles and it has been pointed out in \cite{Hawking} that microcanonical ensemble must be used to discuss thermodynamics of systems where gravitational interactions are important. Extreme cases when such considerations become crucial are quantum black holes which display thermodynamic properties.
\\ \\
In this work, we use microcanonical ensemble to describe confined system of particles in a given background geometry. The main aim is to bring out the purely geometric aspects of phase space volume in curved spacetime, in the hope that these will help understand peculiar thermodynamic properties of systems in a gravitational field. While we do not include self-gravity in our analysis, in the sense that no back reaction on the underlying geometry due to the particles is taken into account, we expect this to be possible based on the results we derive. A large part of the work is devoted to single particle systems which in addition to exhibiting non-trivial and interesting results (like divergence of phase space volume as the system approaches event horizon of a static black hole or as the system size approaches cosmological horizon of a de Sitter spacetime), also forms the basis of statistical mechanical analysis of multi-particle systems. To this end, we have shown that for static spacetimes, the multi-particle phase space volume of ultra-relativistic particles is related to single particle phase space volume via a simple relation which is exactly analogous to the one encountered in classical non-relativistic ideal gases. For generic curved spacetimes (with no assumptions of stationarity etc.), we have obtained leading curvature correction to phase space volumes and entropy of single particle systems using the notion of approximate timelike Killing field to describe an approximately conserved energy. In the case of static spacetimes with this notion of approximately conserved energy, we have shown that the equipartition law of ultra-relativistic particles in flat spacetimes continues to hold.
\\
\\
{\it Conventions}: We work with $(-, +, +, +)$ metric signature. The Greek indices run over spatial components, whereas Latin indices run over both space and time components. Also, we shall be explictly dealing with $(3+1)$ dimensional spacetimes although generalizations to $(D+1)$ dimensional spacetimes will be obvious from our analysis. Finally, the matrix version of a rank-2 tensor $U$ written in some coordinate system will be denoted by $[U]$ and the matrix components as $[U]_{ab}$ where the indices $a,b$ are matrix indices. Also we shall be working in units where $c = k_B = h = 1$.

\section{Constant energy hypersurface: a covariant description}
Let $\{x^i| i=0,1,2,3\}$ be a coordinate chart on the region of interest of some arbitrary spacetime (we are allowing coordinate singularities for this chart). $\{x^\alpha| \alpha=1,2,3\}$ are then a chart on each $x^0 = \text{constant}$ submanifolds. For the discussion below, we assume that our spacetime has a timelike Killing vector field $\xi^i$ and assume no curvature singularities. Good examples would be the exterior of event horizon of a stationary black hole, interior of the cosmological horizon of de Sitter, etc. In Sec-(\ref{Sec4}), we would extend the results to regions of generic curved spacetimes using the notion of approximate Killing vector fields.
\\\\
Suppose we have a container with $N$ free particles in this spacetime. Let $p_{(a)}^i$ be the momentum vector field defined on the world line of the $a$-th particle. The conserved Hamiltonian (Killing energy) of the system is given by\footnote{We assume $H>0$ always.} :
\begin{eqnarray} \label{Hamiltonian}
    H = -\sum_{a=1}^N\xi^ip_{(a)i}
\end{eqnarray}
Now note that defining energy of a particle (or a particle system) in general curved spacetime is non-trivial and highly observer-dependent. In fact if a particle is moving along its spacetime trajectory with momentum $p_i$, then the energy of the particle as measured by an observer with 4-velocity $u^i$ at the location of the particle is given by: 
\begin{equation}
    E = -p_iu^i
\end{equation}
The momentum of the particle on the other hand is obtained from the particle's action in the usual way. Now clearly different observers will measure different energies. But now in our spacetime with a Killing field, we focus on the observers which measure Killing time (i.e. the coordinate time of this class of observers is the Killing time $\tau$). Then, one can write free particle action as
\begin{equation}
    S = \int p_idx^i = \int p_0 d\tau + p_{\mu}\frac{dx^{\mu}}{d\tau}d\tau
\end{equation}
Hence, the Hamiltonian is 
\[
H = -p_0
\]
Now, the Killing field in this coordinate chart is $\xi^i = (1,0,0,0)$ and hence Killing energy is:
\[
H = -\xi^ip_i = -p_0
\]
which matches with free particle energy measured by the Killing observer at the particle's location. An important point to note is that different observers will disagree on the energy of the particle measured by them but they will agree on the Killing energy as it is a space-time scalar. 
\\\\
We would like to re-emphasize that throughout this paper, the word \say{energy} would refer to Killing energy in the context of stationary spacetimes and approximate Killing energy (see Sec-(\ref{Sec4-1})) in general curved spacetimes.
\\\\
Now getting back to our original problem: if the system has at most energy $E$, then phase space volume accessible to the system is given by:
\begin{eqnarray} \label{Theta}
    \Gamma(E) = \frac{1}{N!}\int d^{3N}\mathbf{x}d^{3N}\mathbf{p} \Theta(E-H)
\end{eqnarray}
where we assumed indistinguishable particles. $\Theta$ is the Heaviside step function and integral is over entire phase space. \footnote{The choice of volume form on the one particle phase space to be $d^3\mathbf{x}d^3\mathbf{p} = dx^1 \wedge dx^2 \wedge dx^3 \wedge dp_1 \wedge dp_2 \wedge dp_3$ is actually motivated by the following facts: it is invariant under coordinate transformations in constant $x^0$ foliations. Also in special relativity, it is Lorentz invariant. Showing Lorentz invariance of it is a bit subtle and requires careful analysis; details of which can be found in \cite{VanKampen}. Of course in a general spacetime, under general coordinate transformations that mix space and time it is not an invariant volume form. Also note that simply $d^3\mathbf{x}$ would refer to $dx^1 \wedge dx^2 \wedge dx^3$ and just $d^3\mathbf{p}$ would refer to $dp_1 \wedge dp_2 \wedge dp_3$.} 
\\
\\
Then, accessible phase space volume for the system with energy $E$ is given by:
\begin{eqnarray}
    \Omega(E) = \frac{1}{N!}\int d^{3N}\mathbf{x}d^{3N}\mathbf{p} \hspace{2mm}\delta(E-H)
     = \frac{d\Gamma(E)}{dE}
\end{eqnarray}
Note that the number of accessible microstates is actually defined as:
\begin{eqnarray}
    \Omega(E) = \frac{1}{N! h^{3N}}\int d^{3N}\mathbf{x}d^{3N}\mathbf{p} \hspace{2mm} \delta(E-H) 
\end{eqnarray}
where we have $N$ no. of particles moving in a spacetime with three spatial dimensions and $h$ is Planck's constant. The factor of $h^{3N}$ makes $\Omega(E)$ dimensionless and serves as a quantity denoting denoting \say{smallest phase space volume that can be occupied by a microstate because of Heisenberg's uncertainty}. However, as we have already mentioned, we shall put $h=1$ and assume that things have been made dimensionless.
\\
\\
Then, the microcanonical entropy is defined as:
\begin{eqnarray}
    S = \log \Omega(E)
\end{eqnarray}
and other thermodynamic quantities like temperature, specific heat, etc. can be derived from this.
\subsection{One particle phase space volume}
One can obtain explicit geometric expressions for the above quantities for the phase space of a single particle assuming that the metric $g_{ij}$ can be described in an approximately time independent manner, through (\ref{Hamiltonian}). This is a rather lengthy computation that follows the route similar to the one sketched in \cite{Paddy1}, but generalized to arbitrary metrics; we refer the reader to Appendix-(\ref{Appendix-A}) for details.
\\
\\
One obtains the following expression for the phase space volume of a single particle system of energy at most $E$:
\begin{eqnarray} \label{1P}
    \Gamma(E) = \int_V d^3\mathbf{x}\sqrt{\text{Det}([h])}\bigg(\frac{4\pi}{3}\bigg)\bigg[\lambda + \sum_{\mu}\frac{\gamma_{\mu}^2}{4}\bigg]^{3/2}
\end{eqnarray}
where $V$ is the spatial extent of the box in which the particle resides; the matrix $[h]$ is the matrix form of the metric on spatial foliations defined as:
\begin{eqnarray} \label{1Q}
    h_{\alpha\beta} = g_{\alpha\beta} - \frac{g_{0\alpha}g_{0\beta}}{g_{00}}
\end{eqnarray}
and $\gamma_{\nu} = 2E(g^{0\mu}[\omega]_{\mu\nu})$ and $\lambda = (-g^{00})E^2 - m^2$; further where $[\omega]$ is the matrix that diagonalizes the matrix $[g]$ composed of spatial components of inverse metric to identity matrix. We note that the expression (\ref{1P}) for phase space volume reduces to the expression mentioned in \cite{Paddy1} in case when $g_{0\mu} = 0$.

\subsection{Multi-particle systems} \label{Multi}
For a system with $N$ particles, from (\ref{Theta}) and (\ref{Hamiltonian}), it is obvious that we have to evaluate the integral:
\begin{eqnarray} \label{integral}
    \Gamma(E) = \frac{1}{N!}\int d^3\mathbf{x}d^3\mathbf{p} \hspace{2mm}\Theta\bigg(E-\sum_{k=1}^N H_k\bigg)
\end{eqnarray}
where:
\begin{eqnarray}
    H_k = -\xi^i p_{(k)_i}
\end{eqnarray}
Now, we recall that $\Theta(x)$ has an integral representation given by:
\begin{eqnarray}
    \Theta(x) = \frac{1}{2\pi i} \int_{-\infty}^{\infty} ds \frac{e^{ixs}}{s-i0^+}
\end{eqnarray}
Hence,
\begin{eqnarray} \label{integral1}
    \Theta\bigg(E-\sum_{k=1}^N H_k\bigg) = \frac{1}{2\pi i}\int_{-\infty}^{\infty} ds \frac{e^{iEs}}{s-i0^+}\prod_{k=1}^N e^{-iH_ks}
\end{eqnarray}
Computing this for arbitrary curved spacetime and with non-zero mass is a difficult job. Hence, we focus on $m=0$ case (ultra-relativistic particles or simply massless particles). In the below subsection we compute it for Minkowski spacetime and we shall see that our computation shall give us a direct hint for computing it for some quite general class of spacetimes. 

\subsubsection{Massless (ultra-relativistic) particles in Minkowski}
For Minkowski spacetime, (\ref{integral1}) can be written as:
\begin{eqnarray}
    \Theta\bigg(E-\sum_{k=1}^N H_k\bigg) = \frac{1}{2\pi i}\int_{-\infty}^{\infty} ds \frac{e^{iEs}}{s-i0^+}\prod_{k=1}^N e^{-i(|\mathbf{p_k}|^2 + m^2)^\frac{1}{2}s}
\end{eqnarray}
where $\mathbf{p_k}$ is the 3-momentum of $k$-th particle. 
Thus,
\begin{eqnarray} \label{comp1}
    &&\int \prod_{k=1}^N d^3\mathbf{p_k} \hspace{2mm}\Theta\bigg(E-\sum_{k=1}^N H_k\bigg) \nonumber \\ &=& \frac{1}{2\pi i}\int_{-\infty}^{\infty} ds \frac{e^{iEs}}{s-i0^+}\prod_{k=1}^N \Bigg[d^3\mathbf{p_k} \hspace{2mm}e^{-i(|\mathbf{p_k}|^2 + m^2)^\frac{1}{2}s}\Bigg] \nonumber\\
    &=& \frac{1}{2\pi i}\int_{-\infty}^{\infty} ds \frac{e^{iEs}}{s-i0^+}(4\pi)^N \Bigg[\int_0^{E-Nm} dp \hspace{2mm} p^2 e^{-i(p^2 + m^2)^\frac{1}{2}s}\Bigg]^N
\end{eqnarray}
Note that the limit of integral over $p$ is from $0$ to $E-Nm$ in the last expression as the magnitude of momentum of any particle can never exceed the total energy of the system subtracted by rest mass. In fact any bound larger than or equal to $E-Nm$ shall work. Now, we put $m=0$ and continue the computation of (\ref{comp1}):
\begin{eqnarray} \label{comp2}
        &&\int \prod_{k=1}^N d^3\mathbf{p_k} \hspace{2mm}\Theta\bigg(E-\sum_{k=1}^N H_k\bigg) \nonumber \\ &=& \frac{1}{2\pi i}\int_{-\infty}^{\infty} ds \frac{e^{iEs}}{s-i0^+}(4\pi)^N \Bigg[\int_0^E dp \hspace{2mm} p^2 e^{-ips}\Bigg]^N \nonumber \\
        &=& \frac{(4\pi)^N }{2\pi i}\int_{-\infty}^{\infty} ds \frac{e^{iEs}}{s-i0^+}\Bigg[\frac{2i + e^{-iEs}(-2i + Es(2+iEs))}{s^3}\Bigg]^N
\end{eqnarray}
Now, denoting $f(s) = \frac{2i + e^{-iEs}(-2i + Es(2+iEs))}{s^3}$, it can be easily shown by doing appropriate expansions that $f(s)$ has no pole at $s=0$ in the complex $s$-plane. It's singularity at $s=0$ is a removable one. In fact one obtains:
\begin{eqnarray} \label{comp3}
    \lim_{s \to 0}f(s) = \frac{E^3}{3}
\end{eqnarray}
\\\\
Now, performing the last integral in (\ref{comp2}) by the method of contour integration and closing the contour in the upper half plane (as $E>0$) and noting (\ref{comp3}), we obtain:
\begin{eqnarray}
    \int \prod_{k=1}^N d^3\mathbf{p_k} \hspace{2mm}\Theta\bigg(E-\sum_{k=1}^N H_k\bigg) = \bigg(\frac{4\pi}{3}\bigg)^NE^{3N}
\end{eqnarray}
Considering the system is in a container of volume $V$, we further conclude:
\begin{eqnarray} \label{Result1}
    \Gamma(E) = \frac{1}{N!}\bigg(\frac{4\pi V}{3}\bigg)^NE^{3N}
\end{eqnarray}
Now, note that the integral (\ref{integral}) for massless particles in Minkowski can be written as:
\begin{eqnarray}
        \Gamma(E) &=& \frac{1}{N!}\int dx^\alpha dp_\alpha \Theta\bigg(E-\sum_{k=1}^N H_k\bigg) \nonumber \\ &=& V^N \int \prod_{k=1}^N d^3\mathbf{p_k}\Theta\bigg(E-\sum_{k=1}^N |\mathbf{p_k}|\bigg)\nonumber \\ 
        &=& \frac{1}{N!} (4\pi V)^N \int_0^\infty dp_1 \hspace{1.7mm}... \int_0^\infty dp_N \hspace{3mm} \bigg(\prod_{i=1}^N p_i^2\bigg)\Theta\bigg(E-\sum_{k=1}^N p_k\bigg)
\end{eqnarray}
Thus, comparing this with (\ref{Result1}), we can say that:
\begin{eqnarray} \label{Result2}
    \int_0^\infty dp_1 \hspace{1.7mm}... \int_0^\infty dp_N \hspace{3mm}\bigg(\prod_{i=1}^N p_i^2\bigg) \Theta\bigg(E-\sum_{k=1}^N p_k\bigg) = \bigg(\frac{E^3}{3}\bigg)^N
\end{eqnarray}
We shall be needing this result (\ref{Result2}) in the next subsubsection. 

\subsubsection{Massless (ultra-relativistic) particles in static spacetimes}
For static spacetimes, in addition to having a time-like Killing field, we also have spatial foliations orthogonal to the timelike Killing field such that if $\mu$ denotes spatial components along the foliation, then $g_{0\mu} = 0$. In this case, for $m=0$, the single particle Hamiltonian in (\ref{Ham}) reduces to:
\begin{eqnarray}
    H = \frac{\sqrt{g^{\mu\nu}p_{\mu}p_{\nu}}}{(-g^{00})^\frac{1}{2}}
\end{eqnarray}
Now, using same defintion of $\tilde{p}_{\mu}$ as in (\ref{cond1}), we have:
\begin{eqnarray}
    H = \frac{|\mathbf{\tilde{p}}|}{(-g^{00})^\frac{1}{2}}
\end{eqnarray}
Hence, 
\begin{eqnarray} \label{Equ1}
    \Gamma(E) = \frac{1}{N!}\int \prod_{k=1}^N d^3\mathbf{x_k}\int \prod_{k=1}^N d^3\mathbf{p_k} \bigg(\prod_{i=1}^N \text{Det}([\omega(\mathbf{x_i})]\bigg) \nonumber \\ \Theta\Bigg[E - \bigg(\sum_{i=1}^N \frac{|\mathbf{\tilde{p}_i}|}{\sqrt{-g^{00}(\mathbf{x_i})}}\bigg)\Bigg]
\end{eqnarray}
where as we had defined earlier while doing single particle computation:
\begin{eqnarray}
    [\omega(\mathbf{x_i})]^T[g(\mathbf{x_i})][\omega(\mathbf{x_i})] = [\mathbb{1}]
\end{eqnarray}
So, doing angular integrals in momentum space, we have:
\begin{eqnarray} \label{Equ2}
        \Gamma(E) &=& \frac{(4\pi)^N}{N!}\int \prod_{k=1}^N d^3\mathbf{x_k} \bigg(\prod_{i=1}^N \text{Det}([\omega(\mathbf{x_i})]\bigg) \int_0^\infty dp_1\hspace{1.7mm} ... \int_0^\infty dp_N  \nonumber \\ &&\bigg(\prod_{i=1}^N p_i^2\bigg)\Theta\Bigg[E - \bigg(\sum_{i=1}^N \frac{p_i}{\sqrt{-g^{00}(\mathbf{x_i})}}\bigg)\Bigg] \nonumber
        \\
        &=& \frac{(4\pi)^N}{N!}\int \prod_{k=1}^N d^3\mathbf{x_k} \bigg(\prod_{i=1}^N \text{Det}([\omega(\mathbf{x_i})](-g^{00}(\mathbf{x_i}))^\frac{3}{2}\bigg) \int_0^\infty dq_1\hspace{1.7mm} ... \int_0^\infty dq_N  \nonumber \\ &&\bigg(\prod_{i=1}^N q_i^2\bigg)\Theta\Bigg[E - \sum_{i=1}^N q_i\Bigg] 
\end{eqnarray}
where $q_i = \frac{p_i}{\sqrt{-g^{00}(\mathbf{x_i})}}$
\\
\\
Now, using (\ref{Result2}) in (\ref{Equ2}), we arrive at: 
\begin{eqnarray} \label{Equ3}
    \Gamma(E) = \frac{1}{N!}\bigg(\frac{4\pi E^3}{3}\bigg)^N \int \prod_{i=1}^N \bigg(d^3\mathbf{x_i} \text{Det}([\omega(\mathbf{x_i})])(-g^{00}(\mathbf{x_i}))^{3/2}\bigg)
\end{eqnarray}
Further recall that $\text{Det}([\omega(\mathbf{x_i})]) = \sqrt{\text{Det}([h(\mathbf{x_i})])}$ where in this case $h_{\mu\nu} = g_{\mu\nu} - \frac{g_{0\mu}g_{0\nu}}{g_{00}} = g_{\mu\nu}$. So, denoting the matrix with entries as covariant components of metric ($g_{\mu\nu}$) as $[g_1]$, we have:
\begin{eqnarray} \label{Result3}
    \Gamma(E) = \frac{1}{N!}\bigg(\frac{4\pi}{3}\bigg)^N E^{3N} \Bigg[\int_V d^3\mathbf{x}\bigg(\text{Det}([g_1])(-g^{00})^3\bigg)^{1/2}\Bigg]^N
\end{eqnarray}
Labelling single particle phase volume for energy $E$ as $\Gamma_1(E)$ and fetching its expression from (\ref{1P}) and then comparing with (\ref{Result3}), we conclude that:
\begin{eqnarray} \label{MultiResult}
    \Gamma(E) = \frac{\Gamma_1(E)^N}{N!}
\end{eqnarray}
The result essentially states that for massless particles in spacetimes described by coordinate chart where metric has no spatio-temporal cross components, studying single particle phase volume is enough. The expression (\ref{MultiResult}) is an elementary classical statistical mechanical result for a non-relativistic ideal gas, but here we have shown that this also holds for a relativistic ideal gas in (curved) static spacetimes. 
\\\\
We devote our next sections for studying single particle phase volumes, but however we allow massive particles and most general spacetimes as we have an exact Eq. (\ref{1P}) for single particle system in arbitrary backgrounds.

\section{Single particle system: Exact results for specific spacetimes} \label{Sec3}

\subsection{Uniform Acceleration in Flat Spacetime}
The spacetime metric as described by a particle uniformly accelerated in Cartesian-like coordinate system in a background Minkowski spacetime is given by:
\begin{eqnarray} \label{FNC1}
        g_{00} &=& -(1+a_{\mu}y^{\mu})^2\\
        g_{\mu\nu} &=& \delta_{\mu\nu} 
\end{eqnarray}
where $a^i = u^k\nabla_ku^i$ is the acceleration of the centroid of the box. 
\newline So, for a single free particle in a box, we can obtain $\Gamma(E)$ by performing the integration in the expression (\ref{1P}). Hence, we have:
\begin{eqnarray}
    \Gamma(E) = \int_Vd^3\mathbf{x} \bigg(\frac{4\pi}{3}\bigg)\bigg[\frac{E^2}{(1+a_{\mu}x^{\mu})^2}-m^2\bigg]^{3/2}
\end{eqnarray}
Now, we take the box to be spherical with radius $R$ and denoting magnitude of the 3-acceleration as $|\mathbf{a}|$, we have:
\begin{eqnarray}
    \Gamma(E) = \bigg(\frac{4\pi}{3}\bigg)\int_0^Rdr\hspace{1mm}r^2\int_0^{\pi}d\theta\hspace{1mm}\sin\theta\int_0^{2\pi}d\phi \bigg[\frac{E^2}{(1+|\mathbf{a}|r\cos{\theta})^2}-m^2\bigg]^{3/2}
\end{eqnarray}
Now, the angular integrals for massive case gives hypergeometric functions and then the radial integral is not exactly solvable. So, we rather concentrate on the massless limit. In the massless case, although the integral is exactly solvable, it is still quite complicated. So, we focus on two regimes: firstly, when the box is small $(R << 1/|\mathbf{a}|)$ and when the box approaches the Rindler horizon $(R\rightarrow (1/|\mathbf{a}|)^{-})$
\\
\\
\textbf{Case I: Box radius is small compared to Rindler horizon ${(R << 1/|\mathbf{a}|)}$}
\newline In this regime, we obtain:
\begin{eqnarray} \label{Acc}
    \Gamma(E) &=& \bigg(\frac{8\pi^2}{3}\bigg)\bigg[\frac{2}{3}E^3R^3 + \frac{4}{5}E^3|\mathbf{a}|^2R^5 + \mathcal{O}(R^6)\bigg] \nonumber\\
    &=& \frac{4\pi}{3}E^3.V\bigg[1+ \frac{3}{10\pi}A|\mathbf{a}|^2 + \mathcal{O}(R^3)\bigg]
\end{eqnarray}
where $V$ and $A$ denote volume and area of the spherical box containing the particle.
\\
\\
\textbf{Case II: Box radius approaches Rindler horizon ${(R \rightarrow \left( 1/|\mathbf{a}| \right)^-)}$}
\newline In this regime, the leading order divergent behaviour of $\Gamma(E)$ is captured by the following expression:
\begin{eqnarray} \label{Rinddiv}
    \Gamma(E)_{sing} \sim \bigg(\frac{8\pi^2}{3}\bigg)E^3\Bigg[\frac{1}{2|\mathbf{a}|^4(\frac{1}{|\mathbf{a}|}-R)} - \frac{1}{2}\log\bigg(\frac{1}{|\mathbf{a}|}-R\bigg)\Bigg]
\end{eqnarray}
A similar analysis for accelerated box of classical ideal gas in some symmetric spacetimes with Killing horizons has been done in \cite{SanPad, BhattaPaddy}. In this context, it is worth noting that the leading order divergent behaviour of both phase space volume and entropy, as computed from expression (\ref{Rinddiv}), will be different from the one mentioned in \cite{SanPad, BhattaPaddy}, for two reasons: the difference in spatial geometry of the box (spherical for our case, cubical for \cite{SanPad, BhattaPaddy}), and the ensemble being used (microcanonical vs. canonical).  
\\
\\
\textbf{Note:} The acceleration correction terms in (\ref{Acc}) were area dependent because we took the container to be symmetric, e.g. a spherical box here. However, in general, the correction is not dependent on either the total surface area of the box or the cross-sectional area of the box (cross section perpendicular to the direction of acceleration) as is shown in Appendix-(\ref{AppB1}).

\subsection{Static spherically symmetric black holes}
For a static spherically symmetric black hole, in Schwarzschild coordinates have the line element given by: 
\begin{eqnarray}
    ds^2 = -f(r)dt^2 + \frac{1}{f(r)}dr^2 + r^2d\Omega^2 
\end{eqnarray}
As $g_{0\alpha} = 0$, so $[h] = [g]$ and $\gamma_{\mu} = 0$. Thus, 
\begin{eqnarray} 
        \Gamma(E) &=& \int_V d^3\mathbf{x}\sqrt{\text{Det}([g])}\bigg(\frac{4\pi}{3}\bigg)\lambda^{3/2} \nonumber\\
        &=& \int_V drd\theta d\phi \frac{r^2\sin(\theta)}{\sqrt{f(r)}}\bigg(\frac{4\pi}{3}\bigg)\bigg(\frac{E^2}{f(r)} - m^2\bigg)^{3/2}\label{staticBH}
\end{eqnarray}
From the expression in (\ref{staticBH}) it is readily observed that the phase volume diverges when $f(r)=0$, i.e. for a static black hole when $r=2GM$ (i.e. at the event horizon). So, as the particle container approaches the event horizon, the phase volume that can be occupied by the particle diverges for a given energy $E$. \footnote{Actually, $E$ is the energy of the particle measured at infinity and it is conserved along the geodesic. Also, another thing to note is that the divergence of phase space volume doesn't come only because the locally measured energy is infinitely redshifted $E_{\text{loc}} = E/\sqrt{f}$ at $f=0$, but there is a divergence coming from the volume form itself.} This has also been observed in \cite{Paddy1}. 
\\\\
\textit{\textbf{Computation in Painleve-Gullstrand coordinates and covariance of the phase space volume}}
\\\\
Since Schwarzschild coordinate chart is not well-behaved ar the event horizon, we now redo the computation using Painleve-Gullstrand coordinates which are well-behaved at the horizon. Then in these coordinates, we have the line element given as:
\begin{eqnarray}
    ds^2 = -f(r)dT^2 + 2\sqrt{1-pf(r)}dTdr + pdr^2 + r^2d\Omega^2 
\end{eqnarray}
where $p \in (0,1]$.
\newline It can be obtained that, $g^{00} = -K = -p$. 
So, $\lambda = KE^2 - m^2 = pE^2 - m^2$. 
\newline The $[g]$ matrix for black hole in Painleve coordinates is given by:
\begin{eqnarray}
    [g] = \text{Diag}\hspace{2mm}\bigg(pf(r), \frac{1}{r^2}, \frac{1}{r^2\sin^2\theta}\bigg)
\end{eqnarray}
Clearly $[g]$ is already diagonal and hence:
\begin{eqnarray}
    [\omega] = \text{Diag}\hspace{2mm}\bigg(\frac{1}{\sqrt{pf(r)}}, r, r\sin\theta\bigg)
\end{eqnarray}
Now as $n^{\mu} = g^{0\mu}$, so:
\begin{eqnarray}
    [n]^T = \bigg(\sqrt{1-pf(r}, 0, 0\bigg)
\end{eqnarray}
and as $\gamma_{\nu} = 2E(n^{\mu}[\omega]_{\mu\nu})$, so:
\begin{eqnarray}
    [\gamma]^T = 2E[n]^T[\omega] = \frac{2E}{\sqrt{pf(r)}}\bigg(\sqrt{1-pf(r)}, 0, 0\bigg)
\end{eqnarray}
Hence, putting all together:
\begin{eqnarray} \label{staticBH2}
        \Gamma(E) &=& \int_V drd\theta d\phi \frac{r^2\sin(\theta)}{\sqrt{pf(r)}}\bigg(\frac{4\pi}{3}\bigg)\bigg(pE^2 - m^2 + \frac{E^2(1-pf(r))}{pf(r)}\bigg)^{3/2} \nonumber \\
         &=& \int_V drd\theta d\phi \frac{r^2\sin(\theta)}{\sqrt{pf(r)}}\bigg(\frac{4\pi}{3}\bigg)\bigg(\frac{E^2(1-pf(r)+p^2f(r))}{pf(r)} - m^2\bigg)^{3/2}
\end{eqnarray}
When $p=1$, it can be readily seen that (\ref{staticBH2}) matches with (\ref{staticBH}). Also, note that for any $p \in (0,1]$, (\ref{staticBH2}) diverges when $r=2GM$ is achieved.

\subsection{de Sitter (dS) spacetime}
The dS metric as described in locally inertial coordinates tied to a particle moving along a geodesic is given by \cite{Mashhoon}:
\begin{eqnarray}
    ds^2 = -\cos^2(Hr)dt^2 + dr^2 + \frac{\sin^2(Hr)}{H^2}d\Omega^2
\end{eqnarray}
with $H=\frac{\pi}{2}\sqrt{\Lambda}$ with $\Lambda$ being the cosmological constant.
The curve $r=0$ is the worldline of the particle (or, in our case, the centroid of the box). \footnote{Obviously, $\Lambda = \frac{4H^2}{\pi^2}$ is the cosmological constant and $\frac{1}{\sqrt{\Lambda}} = \frac{\pi}{2H}$ is the radius of cosmological horizon}
It can be easily obtained that:
\begin{eqnarray} \label{dS1}
    \Gamma(E) = \int_V dr d\Omega \bigg(\frac{4\pi}{3}\bigg)\bigg(\frac{E^2}{\cos^2(Hr)} - m^2\bigg)^{\frac{3}{2}}\frac{\sin^2(Hr)}{H^2}
\end{eqnarray}
We assume the box to be a spherical box of radius $R$. So, then after performing angular integrals in (\ref{dS1}), we have:
\begin{eqnarray} \label{dS2}
        \Gamma(E) &=& \frac{16\pi^2}{3}\int_0^Rdr\frac{\sin^2(Hr)}{H^2}\bigg(\frac{E^2}{\cos^2(Hr)} - m^2\bigg)^{\frac{3}{2}} \nonumber \\
        &=& \frac{16\pi^2}{3H^3}\int_0^\Psi dz \sin^2z\bigg(\frac{E^2}{\cos^2z}-m^2\bigg)^{3/2}
\end{eqnarray}
where in the second line we performed change of variables by defining $z = Hr$ and have defined $\Psi = HR$.
\\
\\
The integral in \ref{dS2} is very complicated for general $R$. So, we will check two cases depending on relative value of $R$ w.r.t. cosmological horizon radius $\frac{\pi}{2H}$: firstly, when $R \to 0^+$ (or $R << \frac{\pi}{2H}$) which is equivalently $\Psi \to 0^+$ and secondly, when $R \to \frac{\pi}{2H}^{-}$ which is equivalently $\Psi \to \frac{\pi}{2}^{-}$.
\\
\\
\textbf{Case I :} ${\Psi \to 0^+}; R << \frac{\pi}{2H}$
\\
\newline By performing small $z$ expansion of the integrand of second line in (\ref{dS2}) one can obtain:
\begin{eqnarray} \label{dS3}
        \Gamma(E) &=& \frac{16\pi^2}{3H^3}(E^2 - m^2)^{3/2}\int_0^\Psi dz \bigg(z^2 - \frac{z^4}{3} + \mathcal{O}(z^6)\bigg) \nonumber\\ && \hspace{58mm}\bigg[1+\frac{3}{2}\bigg(\frac{E^2}{E^2-m^2}\bigg)z^2 + \mathcal{O}(z^4)\bigg] \nonumber\\ 
        &=& \frac{16\pi^2}{3H^3}(E^2 - m^2)^{3/2}\int_0^\Psi dz \bigg[z^2+\bigg\{\frac{3}{2}\bigg(\frac{E^2}{E^2-m^2}\bigg) - \frac{1}{3}\bigg\}z^4 + \mathcal{O}(z^6)\bigg] \nonumber\\
        &=& \frac{16\pi^2}{3H^3}(E^2 - m^2)^{3/2}\bigg[\frac{\Psi^3}{3}+\bigg\{\frac{3}{2}\bigg(\frac{E^2}{E^2-m^2}\bigg) - \frac{1}{3}\bigg\}\frac{\Psi^5}{5} + \mathcal{O}(\Psi^7)\bigg]
\end{eqnarray}
Now putting $\psi = HR$ in last expression in (\ref{dS3}), we obtain:
\begin{eqnarray} \label{dSfinal1}
    \Gamma(E) = \frac{16\pi^2}{9}(E^2 - m^2)^{3/2}\bigg[R^3+\bigg\{\frac{9}{2}\bigg(\frac{E^2}{E^2-m^2}\bigg) - 1\bigg\}\frac{H^2R^5}{5} + \nonumber\\ \frac{1}{H^3}\mathcal{O}(H^7R^7)\bigg]
\end{eqnarray}
\\
\\
\textbf{Case II:} ${\Psi \to \frac{\pi}{2}^{-}}; R \to \left(\frac{\pi}{2H}\right)^{-}$
\\
\newline It is obvious that the integral (\ref{dS2}) diverges for $\Psi \to \frac{\pi}{2}$ and unlike the case of static black hole, the divergence is solely from the infinte redshift of energy at the cosmological horizon. The singular part of the integral which diverges at the cosmological horizon can be easily obtained by writing $\sin z = \cos(\frac{\pi}{2} - z)$
and $\cos z = \sin(\frac{\pi}{2} - z)$ and then performing change of variables from $z$ to $x = \frac{\pi}{2} - z$ and then performing a small parameter expansion in x followed by integrating the terms that will result in singular behaviour as $x \to 0$. Hence, one can obtain the singular behaviour of $\Gamma(E)$ as $R \to \frac{\pi}{2H}$ as:
\begin{eqnarray} \label{dSdiv}
    \Gamma(E)_{\text{sing}} = \frac{16\pi^2}{3H^3}\bigg\{\frac{E^3}{(\frac{\pi}{2} - RH)^2}\bigg\} - \bigg\{\frac{1}{2}+\frac{3}{2}\bigg(\frac{m^2}{E^2}\bigg)\bigg\}E^3\ln\bigg(\frac{\pi}{2} - RH \bigg)
\end{eqnarray}
\\
\\
We note that the spatial volume element doesn't contribute to divergence of $\Gamma(E)$ at cosmological horizon for de Sitter and at Rindler horizon for accelerated particle in flat spacetime and hence this divergence is somewhat different from the divergence of $\Gamma(E)$ at event horizon for Schwarzschild black hole where both the spatial volume element and the $\lambda$ factor ($-g^{00}$ term) with energy ($E$) diverges at event horizon.

\section{Leading curvature corrections in arbitrary curved spacetimes}\label{Sec4}
\subsection{Energy in arbitrary curved spacetimes} \label{Sec4-1}
Note that on a general curved spacetime, there will be no Killing field and hence no notion of conserved energy. Without the notion of a Killing field (and hence of a conserved energy), the whole concept of thermal equilibrium doesn't make any sense. However, one can consider approximate Killing fields defined near a point on the trajectory of the centroid of the box in a patch of spacetime which is small enough compared to variation of curvature tensor spatially as well as temporal variation of curvature due to the particle's motion. We will make this notion more precise in the following paragraph by introducing Fermi Normal Coordinates (FNC) and performing computations in that chart.
\\\\
The FNC constitute a coordinate system on spacetime which is constructed around a timelike curve in spacetime (which is the trajectory of the centroid of the box) such that the coordinate chart is flat (metric is Minkowski and Christoffel symbols vanish) on the curve. The coordinate chart is given by $\{\tau, y^1, y^2, y^3\}$ where $\tau$ is proper time along the curve. The metric components in FNC is given in \cite{DK} to be:
\begin{eqnarray} \label{FNC}
        g_{00} = -(1+a_{\mu}y^{\mu})^2 + R_{0\mu0\nu}y^{\mu}y^{\nu} + \mathcal{O}(y^3, \partial R) \\
        g_{0\mu}= -\frac{2}{3}R_{0\rho\mu\nu}y^{\rho}y^{\nu} + \mathcal{O}(y^3, \partial R) \\
        g_{\mu\nu} = \delta_{\mu\nu} - \frac{1}{3}R_{\mu\rho\nu\sigma}y^{\rho}y^{\sigma} + \mathcal{O}(y^3, \partial R)
\end{eqnarray}
where $a^i = u^k\nabla_ku^i$ is the acceleration of the centroid of the box. Now consider the vector field $\xi^i = (1,0,0,0)$ in this chart, which coincides with the four velocity of the box centroid on the trajectory. Since we will define Hamiltonian using evolution generated by $\xi^i$, we will first briefly indicate the extent to which $\xi^i$ might be considered as an approximate Killing field, so that the corresponding energy can be taken as approximately constant. To do so, we follow a variant of the analysis given in Appendix A.1 of \cite{DawoodThermo} for approximate boost Killing fields. For the present case, the computation is given in Appendix-(\ref{Appendix-C.1}). It follows from this analysis that for constant acceleration $\dot{a}^i = 0$ (where dot represents derivative w.r.t. $\tau$): 
\begin{eqnarray} \label{KVF1}
    \nabla_i\xi_j + \nabla_j\xi_i = \mathcal{O}(y^2)
\end{eqnarray}
and for $\dot{a}^i \neq 0$:
\begin{eqnarray} \label{KVF2}
    \nabla_i\xi_j + \nabla_j\xi_i = \mathcal{O}(y)
\end{eqnarray}
Hence this approximate Killing field (it is approximate as it satisfies Killing equation only up to an order in $y$ for non-constant acceleration and up to an order in $y^2$ for constant acceleration) helps us to define an \say{approximately conserved} Killing energy which is defined for particles confined in a box with spatial extent smaller than the length scale of spatial variation of curvature and conserved for time scales smaller compared to the temporal variation of the curvature tensor along the trajectory of the centroid of the box. We discuss a little bit more on this with an example in Appendix-(\ref{C.2}).

\subsection{One particle phase space volume}
In the earlier sections, we performed analysis in some special spacetimes. Now, we do a perturbative analysis on any general spacetime for particles confined in a box small enough compared to curvature scale of the spacetime. Hence, around the worldline of centroid of the box, we can perform a perturbative analysis using Fermi normal coordinates. To this end, consider a perturbed metric over Minkowski where the metric tensor is given by:
\begin{eqnarray}
    g_{ab} = \begin{bmatrix}
        -1-2\epsilon q & \epsilon l_1 & \epsilon l_2 & \epsilon l_3 \\
        \epsilon l_1 & 1+\epsilon h_{11} & \epsilon h_{12} & \epsilon h_{13} \\
        \epsilon l_2 & \epsilon h_{12} & 1+\epsilon h_{22} & \epsilon h_{23} \\
        \epsilon l_3 & \epsilon h_{13} & \epsilon h_{23} & 1+\epsilon h_{33}
    \end{bmatrix}
\end{eqnarray}
From (\ref{spatialmetric}), we obtain:
\begin{eqnarray}
    h_{\alpha\beta} = g_{\alpha\beta} + \mathcal{O}(\epsilon^2)
\end{eqnarray}
Hence, 
\begin{eqnarray} \label{Detresult}
    \sqrt{\text{Det}([h])} &=& [1+\epsilon(h_{11}+h_{22}+h_{33})+\mathcal{O}(\epsilon^2)]^{1/2} \nonumber \\ &=& 1+\epsilon\bigg(\frac{h_{11}+h_{22}+h_{33}}{2}\bigg)+\mathcal{O}(\epsilon^2)
\end{eqnarray}
Also, we have:
\begin{eqnarray}
    -g^{00} = K = 1-2\epsilon q + \mathcal{O}(\epsilon^2)
\end{eqnarray}
Now, note that as $\gamma_{\nu} = n^{\mu}\omega_{\mu\nu}$ and as $n^{\mu} = g^{0\mu} = \epsilon l^{\mu}$, so, $\gamma_{\nu}^2 \sim \mathcal{O}(\epsilon^2)$. 
Thus volume occupied by constant energy hypersurface in 3-momentum space is given using (\ref{Master}):
\begin{eqnarray}
    \int d^3\mathbf{p} \hspace{2mm} \Theta(E+\xi^ip_i) = \bigg(\frac{4\pi}{3}\bigg)\bigg[(1-2\epsilon q)E^2-m^2 + \mathcal{O}(\epsilon^2)\bigg]^{3/2} \nonumber \\ \bigg[1+\frac{\epsilon}{2}(h_{11}+h_{22}+h_{33}) + \mathcal{O}(\epsilon^2)\bigg]
\end{eqnarray}
This can be simplified to finally obtain:
\begin{eqnarray} \label{pertmaster}
    \int d^3\mathbf{p} \hspace{2mm}\Theta(E+\xi^ip_i) = \bigg(\frac{4\pi}{3}\bigg)(E^2-m^2)^{3/2}\bigg[1+\epsilon\bigg\{\bigg(\frac{h_{11}+h_{22}+h_{33}}{2}\bigg) \nonumber \\ - 3\epsilon q\bigg(\frac{E^2}{E^2-m^2}\bigg)\bigg\} + \mathcal{O}(\epsilon^2)\bigg]
\end{eqnarray}


\noindent We can now proceed to evaluate the curvature corrections by inserting the explicit forms for $q$ and $h_{\mu \nu}$ in Fermi normal coordinates (FNC) as we introduced in (\ref{FNC})
\footnote{
Our convention will be as follows: Repeated indices when one is in upstairs and other is in downstairs will be automatically summed unless otherwise stated. But for summing over repeated indices placed at the same level, an explicit summation symbol will be used.    
}. 
\\
\\
Now, we shall assume that our box is a spherical box. \footnote{This is motivated by the fact that we shall be using our calculations mostly for spherically symmetric spacetimes and a spherical box is a natural choice for it} This will require a coordinate change as FNC as described in (\ref{FNC}) are adapted to a locally Cartesian-like coordinate chart. We shall not write FNC in locally spherical coordinate chart, rather we shall do change of variables in the integration. 
\\
\\
Now, to proceed, we shall first consider curvature as the small parameter (note that $\mathcal{O}(y^3)$ terms in FNC also contain higher order curvature terms, so, we can equivalently treat the box radius as well as curvature as small parameters) and then invert the metric. Then, we will consider acceleration to be small and keep only terms till quadratic order in acceleration. Note that dimensionally, acceleration is $[L]^{-1}$, whereas curvature is $[L]^{-2}$ and hence we keep linear order terms in curvature and quadratic order in acceleration. So, essentially, the dimensionless small parameters in our perturbative calculations are essentially $R^2.(\text{curvature})$ and $R^2|\mathbf{a}|^2$ and we consider both to be of same order. So, we obtain the following till first order in curvature and quadratic order in acceleration. 
\begin{eqnarray}
    K = -g^{00} = 1-2a_{\mu}y^{\mu} + (R_{0\mu 0\nu} + 3a_{\mu}a_{\nu})y^{\mu}y^{\nu} \\
    h_{\mu\nu} = g_{\mu\nu} = \delta_{\mu\nu} - \frac{1}{3}R_{\mu\rho\nu\sigma}y^{\rho}y^{\sigma}
\end{eqnarray}
where $h_{\alpha\beta}$ is the metric on spatial foliations as appearing in (\ref{1P}). Now, using (\ref{Detresult}), we say that:
\begin{eqnarray}
    \sqrt{\text{Det}([h])} = 1 -\frac{1}{6}\sum_{\mu=1}^3R_{\mu\alpha\mu\beta}y^{\alpha}y^{\beta}
\end{eqnarray}
Also, spatio-temporal cross terms of the metric appear only in quadratic order in (\ref{1P}) and hence quadratic in curvature. So, we ignore them in our computation. 
\newline So, putting all of them in (\ref{1P}) retaining only the order of terms we are interested in (first order in curvature and second order in acceleration) and also ignoring all acceleration-curvature cross terms appearing in $\mathcal{O}(y^3)$, the phase volume is:
\begin{eqnarray} \label{MasterEqn}
    \Gamma(E) = \int_Vd^3\mathbf{y}\bigg(\frac{4\pi}{3}\bigg)\bigg(1 -\frac{1}{6}\sum_{\mu=1}^3R_{\mu\alpha\mu\beta}y^{\alpha}y^{\beta}\bigg)\bigg[(1-2a_{\mu}y^{\mu} + \nonumber \\ (R_{0\mu 0\nu} + 3a_{\mu}a_{\nu})y^{\mu}y^{\nu})E^2-m^2\bigg]^{3/2} \nonumber \\
    = \int_Vd^3\mathbf{y}\bigg(\frac{4\pi}{3}\bigg)(E^2-m^2)^{3/2}\bigg[1-\frac{1}{6}\sum_{\mu=1}^3R_{\mu\alpha\mu\beta}y^{\alpha}y^{\beta}+ \nonumber \\\bigg(\frac{E^2}{E^2-m^2}\bigg)\bigg\{-3a_{\mu}y^{\mu}+\frac{3}{2}(R_{0\mu 0\nu} + 3a_{\mu}a_{\nu})y^{\mu}y^{\nu}\bigg\} + \frac{3}{2}\bigg(\frac{E^2}{E^2-m^2}\bigg)^2a_{\mu}a_{\nu}y^{\mu}y^{\nu}\bigg]
\end{eqnarray}
where the spatial integrations are to be performed inside a spherical box of radius $R$.  
\newline Replacing Cartesian $\{y^1,y^2,y^3\}$ with the usual spherical polar $\{r, \theta, \phi\}$ by the usual coordinate transformation, one can write: \footnote{And also note the facts that $\int dr\hspace{1mm} r^2 d\Omega_2 \hspace{1mm}x$ = $\int dr \hspace{1mm} r^2 d\Omega_2 \hspace{1mm}y$ = $\int dr \hspace{1mm}  r^2 d\Omega_2 \hspace{1mm}z$ = 0 and also $\int dr \hspace{1mm} r^2 d\Omega_2 \hspace{1mm}xy$ = $\int dr \hspace{1mm} r^2 d\Omega_2 \hspace{1mm} yz$ = $\int dr\hspace{1mm}  r^2 d\Omega_2 \hspace{1mm} zx$ = 0}
\begin{eqnarray} \label{eq1}
    && \int_V d^3 {\bf y} \bigg(-\frac{1}{6}\sum_{\mu=1}^3R_{\mu\alpha\mu\beta}y^{\alpha}y^{\beta}\bigg)\nonumber \\ &=& -\frac{1}{6}\sum_{\mu = 1}^3\int_V dr d\Omega r^2 R_{\mu\alpha\mu\beta}y^{\alpha}y^{\beta} \nonumber \\ &=& \frac{4\pi R^5}{45}\tensor{G}{^0_0}
\end{eqnarray}
where $V$ denotes a spherical box at centered at origin with radius $R$, and $d\Omega=\sin{\theta} d\theta d\phi$. Note that in evaluating the above integral (\ref{eq1}), we have used the fact that:
\begin{eqnarray}
    \int_V dr r^2 d\Omega_2 x^2 = \int_V dr r^2 d\Omega_2 y^2 = \int_V dr r^2 d\Omega_2 z^2 = \bigg(\frac{4\pi}{3}\bigg)\frac{R^5}{5}
\end{eqnarray}
and \footnote{To prove the following (\ref{geom1}), first note that $R_{\mu\nu\mu\nu} = \tensor{R}{^\mu^\nu_\mu_\nu}$ (no summation over any index) as the metric is Minkowski along the curve and as these are Riemann tensors defined on the curve, so the indices can be freely raised or lowered. Then, $\sum_{\mu\nu}R_{\mu\nu\mu\nu} = \tensor{R}{^\mu^\nu_\mu_\nu}$. Finally note that $\tensor{R}{^1^2_1_2}+\tensor{R}{^2^3_2_3}+\tensor{R}{^1^3_1_3}=-\tensor{G}{^0_0}$}
\begin{eqnarray}\label{geom1}
    \sum_{\mu\nu}R_{\mu\nu\mu\nu} = -2\tensor{G}{^0_0}
\end{eqnarray}
Further we note that:
\begin{eqnarray}
    \int_V d^3 {\bf y} \hspace{2mm}y^{\mu} = 0    
\end{eqnarray}
and 
\begin{eqnarray}
    \int_V d^3 {\bf y} \hspace{2mm}R_{0\mu 0\nu}y^{\mu}y^{\nu} = \bigg(\frac{4\pi}{3}\bigg)\frac{R^5}{5}\sum_{\mu}R_{0\mu 0\mu} = \bigg(\frac{4\pi}{3}\bigg)\frac{R^5}{5}\tensor{R}{^0_0}
\end{eqnarray}
and hence similarly,
\begin{eqnarray}
    \int_Vd^3\mathbf{y} \hspace{2mm}a_{\mu}a_{\nu}y^{\mu}y^{\nu} = \bigg(\frac{4\pi}{3}\bigg)\frac{R^5}{5}\sum_{\mu}(a_{\mu})^2 = \bigg(\frac{4\pi}{3}\bigg)\frac{R^5}{5}|\mathbf{a}|^2
\end{eqnarray}
Putting all of them in (\ref{MasterEqn}), we obtain:
\begin{eqnarray} \label{big}
    \Gamma(E) = \frac{4\pi}{3}(E^2-m^2)^{3/2}\bigg[\frac{4\pi}{3}R^3 + \frac{4\pi R^5}{45}\tensor{G}{^0_0} + \nonumber\\ \bigg(\frac{E^2}{E^2-m^2}\bigg)\bigg(\frac{2\pi R^5}{5}\tensor{R}{^0_0} + \frac{6\pi R^5}{5}|\mathbf{a}|^2\bigg) + \bigg(\frac{E^2}{E^2-m^2}\bigg)^2\bigg(\frac{2\pi R^5}{5}|\mathbf{a}|^2\bigg)\bigg]
\end{eqnarray}
Now, writing in terms of volume and area of the box and clubbing curvature dependent terms and acceleration dependent terms separately together, we obtain:
\begin{eqnarray} \label{Big}
    \Gamma(E) = \frac{4\pi}{3}(E^2-m^2)^{3/2}.V\bigg[1+ \frac{A}{60\pi}\tensor{G}{^0_0} + \bigg(\frac{E^2}{E^2-m^2}\bigg)\frac{3A}{40\pi}\tensor{R}{^0_0} + \nonumber\\ \bigg\{\bigg(\frac{E^2}{E^2-m^2}\bigg)\frac{9}{40\pi} +\bigg(\frac{E^2}{E^2-m^2}\bigg)^2\frac{3}{40\pi}\bigg\}A|\mathbf{a}|^2\bigg]
\end{eqnarray}
We again emphasize that (\ref{Big}) is obtained by taking first order in curvature and quadratic order in acceleration terms. Also, we emphasize the point that Eq. (\ref{Big}) is only valid when $V$ and $A$ are computed for a spherical box and it is not valid for boxes of arbitrary geometry.
\\
\\
We can perform two important consistency checks:
\begin{itemize}
    \item By putting dS values of $\tensor{R}{^0_0} = 3H^2$ and $\tensor{G}{^0_0} = -3H^2$ and putting $a^i = 0$ in (\ref{big}), one gets back the small $R$ expression (\ref{dSfinal1}) for $\Gamma(E)$ in dS. 
    \item By putting mass and curvature terms to zero, we get back Rindler result (\ref{Acc}) when box radius is very small compared to Rindler horizon.
\end{itemize}

\noindent\textbf{Note:} The first order curvature correction terms were area-dependent and depended only on the time-time components of Ricci and Einstein tensors in (\ref{Big}) because we took our system to be symmetrical (here a spherical box). But, for an asymmetric container, like a cuboidal box, the correction terms are neither area dependent nor does it depend only on the time-time components of Ricci and Einstein tensors. The expression for a cuboidal box is derived in Appendix-(\ref{AppB2}).

\subsection{Microcanonical entropy and temperature} \label{FNCM}
We have already noted throughout our paper that single-particle statistical mechanics in non-trivial backgrounds is both interesting (e.g. shows some peculiar properties in some symmetric stationary spacetimes, see Sec.-(\ref{Sec3})) as well as is useful for understanding multi-particle systems (e.g. for massless particles in static spacetime one can obtain multi-particle phase volume directly from single particle result via (\ref{MultiResult})). Hence in this subsection, we shall push single particle computations further and understand microcanonical entropy and temperature for such systems.
\\
\\
The number of accessible microstates for a given value of energy $E$ can be obtained from (\ref{Big}) as:
\begin{eqnarray} \label{microstates}
        \Omega(E) &=& \frac{d\Gamma(E)}{dE} \nonumber\\
        &=& 4\pi (E^2-m^2)^{\frac{1}{2}}E \; V \biggl[1+\frac{A}{60\pi}\tensor{G}{^0_0}+ \frac{A}{40\pi}\bigg(\frac{3E^2-2m^2}{E^2-m^2}\bigg)\tensor{R}{^0_0} + \nonumber \\ &&\frac{A}{40\pi}\bigg(\frac{12E^4-19E^2m^2+6m^4}{(E^2-m^2)^2}\bigg)|\mathbf{a}|^2\biggl]
\end{eqnarray}
Now, taking its logarithm and using the fact that $R^2|\mathbf{a}|^2 \sim R^2(\text{curvature}) << 1$, the microcanonical entropy is given by:
\begin{eqnarray} \label{entropy1}
        S(E) &=& \log{\Omega(E)} 
        \nonumber\\
        &=& \log\{4\pi E(E^2-m^2)^{1/2}\} + \log V + A\Bigg[\frac{1}{60\pi}\tensor{G}{^0_0}+ \nonumber\\&& \hspace{-7mm}\frac{1}{40\pi}\bigg(\frac{3E^2-2m^2}{E^2-m^2}\bigg)\tensor{R}{^0_0} +\frac{1}{40\pi}\bigg(\frac{12E^4-19E^2m^2+6m^4}{(E^2-m^2)^2}\bigg)|\mathbf{a}|^2\Bigg]
\end{eqnarray}
Now, the inverse temperature $\beta$ is given by:
\begin{eqnarray} \label{Temperature}
        \beta &=& \frac{\partial S(E)}{\partial E} \nonumber \\ 
        &\approx& \frac{1}{E} + \frac{E}{E^2-m^2} - A\Bigg[\frac{1}{20\pi}\bigg(\frac{Em^2}{(E^2-m^2)^2}\bigg)\tensor{R}{^0_0} + \nonumber \\ &&\frac{1}{20\pi}\bigg(\frac{Em^2(5E^2-7m^2)}{(E^2-m^2)^3}\bigg)|\mathbf{a}|^2\Bigg]
\end{eqnarray}
From Eq. (\ref{entropy1}) and (\ref{Temperature}) it is obvious that leading order curvature correction terms in entropy and inverse temperature are proportional to surface area of the system.

\subsection{Multi-particle systems}
In Section-(\ref{Multi}), we discussed about phase volume occupied by massless particles in static spacetimes. In this restricted case, we had obtained (\ref{Result3}) and hence subsequently (\ref{MultiResult}). Now, we can employ them and use the one particle computations in Section-(\ref{FNCM}) to derive microcanonical entropy and temperature expressions for these $N$-particle systems in a general static spacetime.
\\
\\
From (\ref{MultiResult}), we have:
\begin{eqnarray}
    \Omega(E) &=& \frac{1}{(N-1)!}\Gamma_1(E)^{N-1}\frac{d\Gamma_1(E)}{dE} \nonumber \\
    &=& \frac{1}{(N-1)!}\Gamma_1(E)^{N-1}\Omega_1(E)
\end{eqnarray}
where $\Omega_1(E)$ is single particle constant energy phase volume. 
\newline Hence, entropy will be:
\begin{eqnarray}
    S(E) &=& \log\Omega(E) \nonumber \\
    &=& -\log((N-1)!) + (N-1)\log(\Gamma_1(E))+ \log\Omega_1(E) 
\end{eqnarray}
Now, employing Stirling's approximation and taking thermodynamic limit (N very large), we have:
\begin{eqnarray} \label{Nparticle}
    S(E) = N - N\log N + N \log(\Gamma_1(E)) + \log(\Omega_1(E))
\end{eqnarray}
Now, we can employ this to evaluate entropy of massless limits of many systems of interest. However, we focus our attention to an accelerated system in curved spacetime with the assumption $R^2(\text{curvature}) \sim R^2|\mathbf{a}|^2 << 1$.
\\
\\
Taking massless cases of Eq. (\ref{entropy1}), (\ref{Big}) and plugging them in Eq. (\ref{Nparticle}), we get:
\begin{eqnarray} \label{BigFormula}
    S(E) = N - N\log N + N\bigg[\log\bigg(\frac{4\pi}{3}E^3\bigg) + \log V + A\bigg(\frac{1}{60\pi}\tensor{G}{^0_0} +\nonumber \\ \frac{3}{40\pi}\tensor{R}{^0_0} + \frac{3}{10\pi}|\mathbf{a}|^2\bigg)\bigg] + \log\{4\pi E^2\} + \log V + A\bigg[\frac{1}{60\pi}\tensor{G}{^0_0}+ \nonumber \\ \frac{3}{40\pi}\tensor{R}{^0_0} +\frac{3}{10\pi} |\mathbf{a}|^2\bigg]
\end{eqnarray}
Now, clubbing all $N$ dependent (and $E, V, A$, curvature and acceleration independent) constants as $C_N$ and taking thermodynamic limit (large N limit), we can further simplify the expression (\ref{BigFormula}) to obtain:
\begin{eqnarray} \label{NiceResultNparticles1}
    S(E) \approx C_N + 3N\log E + N\log V + N \; A \; \bigg[\frac{1}{60\pi}\tensor{G}{^0_0}+ \bigg(\frac{3}{40\pi}\bigg)(\tensor{R}{^0_0} + 4 |\mathbf{a}|^2)\bigg]
\end{eqnarray}
Furthermore, we note that the microcanonical inverse temperature $\beta$ is given by:
\begin{eqnarray} \label{NiceResultNparticles2}
    \beta = \frac{3N}{E} 
\end{eqnarray}
So, for massless particles in static spacetimes, entropy has leading order curvature correction correction terms appearing as proportional to area in Eq. (\ref{NiceResultNparticles1}), but the inverse temperature has no first order curvature (or quadratic order acceleration) correction terms as is evident from Eq. (\ref{NiceResultNparticles2}) \footnote{Note that one particle $\beta$ also has no first order curvature correction terms if particles are massless even in a completely arbitrary spacetime as is evident by putting $m=0$ in (\ref{Temperature})}. Here we note that it has been established a long time back \cite{Tolman} that the general expression for equipartition of energy in special relativity for a single particle is:
\begin{eqnarray}
    \Big\langle \frac{p_{\mu}^2}{E}\Big\rangle = T
\end{eqnarray}
with $\langle\cdot\rangle$ denoting thermal average. Hence with $\mu = 1,2,3$ and in massless limit:
\begin{eqnarray} \label{Tolman}
    \langle E\rangle = 3T
\end{eqnarray}
In fact our result (\ref{NiceResultNparticles2}) shows that this equipartition expression (\ref{Tolman}) continues to hold even for curved static spacetimes where approximately conserved energy is appropriately defined.

\section{Discussion}
To summarize, in this work we have explicitly computed correction terms to microcanonical phase volume, entropy and temperature of inertial and non-inertial systems in curved stationary spacetimes (so that we have a notion of a \say{conserved energy}). We have further observed that these corrections to thermodynamic quantities lead to quite non-trivial consequences such as divergent behavior near null surfaces as well as area dependent curvature and acceleration terms. We highlight some of the major qualitative inferences of this work in the following paragraphs.
\\
\\
The divergent behavior of $\Gamma(E)$ (and hence other thermodynamic quantities, including entropy) at event horizon of Schwarzschild black hole (see (\ref{staticBH2})) is due to infinite redshift of energy (i.e. divergence of $(-g^{00})$ term) as well as due to divergence of spatial volume element. However, the divergence of $\Gamma(E)$ at cosmological horizon of de Sitter (see (\ref{dSdiv})) and Rindler horizon for uniform acceleration in flat spacetime (see (\ref{Rinddiv})) is only due to infinite redshift of energy. 
\\
\\
Apart from that, it is also worth pointing out that although quite appealing, the area dependent curvature correction terms appearing in entropy expressions of (\ref{entropy1}) and (\ref{NiceResultNparticles1}) have no direct relationship with the Bekenstein-Hawking entropy of black holes. This is because we are not considering self-gravitation here and hence the system does not describe a gravitational collapse or black hole. Nevertheless, we expect that the above analysis would still hold provided one uses, for $G_{ab}$ and $R_{ab}$, their expressions as obtained from Einstein equations with the system itself as a source. This work is under progress.
\\
\\
We also note that entropy of massless $N$-particle systems for accelerated motion in curved static spacetime has area dependent first order correction terms. But, microcanonical inverse temperature has no correction term at this order and just reproduces the standard  equi-partition result for ultra-relativistic ideal gas in flat background. 
\\
\\
As already emphasized, the most natural and relevant extension of this work is to include self-gravitation of the system itself into account (through the metric), and compute the various thermodynamic quantities. This would be the first step towards understanding the purely statistical mechanical contribution to thermodynamic properties of black holes. We would also like to point out that beyond point particle systems, there has been studies on thermodynamics in curved backgrounds for thermal field theories. For example, the effects of spacetime curvature for self-interacting statistical field theories has been studied in \cite{NathBhatta} where curvature contributions to spontaneous symmetry breaking and associated curvature correction terms to critical temperature has been discussed; ignoring back-reaction of the field on the metric (in the same spirit to our current work). Although relating the present work on free point particle thermodynamics with the work \cite{NathBhatta} is beyond the scope of this paper, it would worth clarifying the connection.

\appendix
\section{Derivation of the Formula for ${\Gamma(E)}$} \label{Appendix-A}
Consider a metric $g_{ij}$ on a stationary spacetime, in a chart with time coordinate $x^0$. In this chart, the timelike Killing field is $\xi^i = (1,0,0,0)$.
So if $p_i$ is the free particle conjugate momentum,
\begin{eqnarray}
    H = -\xi^ip_i = -p_0
\end{eqnarray}
and it is conserved if the particle follows a geodesic trajectory. From the mass shell condition $p_ip^i = -m^2$, we obtain
\begin{eqnarray}
   K(p_0)^2 - (2n^{\mu}p_{\mu})p_0 -(g^{\mu\nu}p_{\mu}p_{\nu} + m^2) = 0 \label{Quad1}
\end{eqnarray}
where $g^{00} = -K$ and $g^{0\mu} = n^{\mu}$. This quadratic equation (\ref{Quad1}) has solutions:
\begin{eqnarray} \label{Sol1}
    p_0^{\pm} = \frac{n^{\mu}p_{\mu} \pm \sqrt{(n^{\mu}p_{\mu})^2 + K(g^{\mu\nu}p_{\mu}p_{\nu} + m^2)}}{K}
\end{eqnarray}
Since we need $H = -p_0 >0$, hence we take $p_0 < 0$ solution out of (\ref{Sol1}) and assuming that $K>0$ and $\mathbf{p}.\mathbf{p}= g^{\mu\nu}p_{\mu}p_{\nu} > 0$ we take the $p_0^{-}$ solution. Thus,
\begin{eqnarray} \label{Ham}
    H = \frac{-n^{\mu}p_{\mu} + \sqrt{(n^{\mu}p_{\mu})^2 + K(g^{\mu\nu}p_{\mu}p_{\nu} + m^2)}}{K}
\end{eqnarray}
To evaluate the integral (\ref{Theta}), we first evaluate $\int dp_\alpha \Theta(E+\xi^ip_i)$ which essentially corresponds to the volume occupied by the surface $E = H$ in the 3-momentum space. 
\newline The surface $H = E$ in the 3-momentum space is given by:
\begin{eqnarray}
    &  (KE + n^{\mu}p_{\mu})^2 = (n^{\mu}p_{\mu})^2 + K(g^{\mu\nu}p_{\mu}p_{\nu} + m^2) \nonumber \\
    & \hspace{-12mm}\implies g^{\mu\nu}p_{\mu}p_{\nu} - 2En^{\mu}p_{\mu} = KE^2 - m^2 \label{Surf1}
\end{eqnarray}
Now for the following part of the computation of volume occupied by this surface, objects with square brackets are matrices on the 3-momentum space (assumed as a vector space) and their indices are not tensor indices but matrix indices and always placed downstairs. However Einstein summation convention will still be in force in the sense that repeated indices will be summed but placement of indices won't matter. Now, define:
\begin{eqnarray} \label{cond1}
    p_{\mu} = [\omega]_{\mu\nu}\tilde{p}_{\nu} \hspace{4mm} \text{s.t.} \hspace{2mm} g^{\mu\nu}p_{\mu}p_{\nu} = [\mathbb{1}]_{\mu\nu}\tilde{p}_{\mu}\tilde{p}_{\nu}
\end{eqnarray}
So, the surface (\ref{Surf1}) can be written in $\title{p}$ coordinate system in momentum space as:
\begin{eqnarray}
        & \sum_{\mu}(\tilde{p}_{\mu})^2 - 2E(n^{\mu}[\omega]_{\mu\nu})\tilde{p}_{\nu} = KE^2 - m^2 \\
        & \hspace{-12mm}\implies \sum_{\mu}(\tilde{p}_{\mu}^2 - \gamma_{\mu}\tilde{p}_{\mu}) = \lambda  \label{Surf2}
\end{eqnarray}
where $\gamma_{\nu} = 2E(n^{\mu}[\omega]_{\mu\nu})$ and $\lambda = KE^2 - m^2$.
\newline Now, completing squares in (\ref{Surf2}), we arrive at the equation of the surface to be: 
\begin{eqnarray} \label{Surf3}
    \sum_{\mu}\bigg(\tilde{p}_{\mu} - \frac{\gamma_{\mu}}{2}\bigg)^2 = \lambda + \sum_{\mu}\frac{\gamma_{\mu}^2}{4}
\end{eqnarray}
Now, clearly the volume occupied by surface (\ref{Surf3}) in 3-momentum space (which essentially a sphere in $\tilde{p}$ coordinate system) and hence the momentum integral that we set forth to evaluate first is given by:
\begin{eqnarray} \label{Master}
    \int d^3\mathbf{p} \Theta(E+\xi^ip_i) = \frac{4}{3}\pi\bigg[\lambda + \sum_{\mu}\frac{\gamma_{\mu}^2}{4}\bigg]^{3/2}\text{Det}([\omega])
\end{eqnarray}
where the factor of Det$([\omega])$ comes because of change of variables in integration. 
\\
\\
Now, the change of basis condition (\ref{cond1}) can be written in matrix notation as:
\begin{eqnarray} \label{cond2}
    [p] = [\omega][\tilde{p}] \hspace{4mm} \text{s.t.} \hspace{2mm} [p]^T[g][p] = [\tilde{p}]^T[\tilde{p}]
\end{eqnarray}
where [p] is column vector corresponding to the 3-momentum vector $p^{\alpha}$ and [g] is the matrix corresponding to $g^{\mu\nu}$. 
So, (\ref{cond2}) implies that:
\begin{eqnarray}
         &[\tilde{p}]^T[\omega]^T[g][\omega] [\tilde{p}] =  [\tilde{p}]^T [\tilde{p}] \\
         &\hspace{-12mm}\implies [\omega]^T[g][\omega] = [\mathbb{1}]
\end{eqnarray}
So, we have:
\begin{eqnarray} \label{Omega}
    [\omega] = [D][\chi]
\end{eqnarray}
where [D] is the matrix with eigenvectors of [g] as columns \footnote{[g] being symmetric, it is orthogonally diagonalizable and hence $[D]^T = [D]^{-1}$} and 
\begin{eqnarray}
    [\chi] = \text{Diag}\bigg(\frac{1}{\sqrt{\lambda_1}}, \frac{1}{\sqrt{\lambda_2}}, \frac{1}{\sqrt{\lambda_3}}\bigg)
\end{eqnarray}
where $\lambda_i$'s are eigenvalues of [g]. 
\\
\\
In \cite{Landau}, it has been shown that the spatial line element on constant time foliations is given by:
\begin{eqnarray}
    dl^2 = h_{\alpha\beta}dx^{\alpha}dx^{\beta}
\end{eqnarray}
where
\begin{eqnarray} \label{spatialmetric}
    h_{\alpha\beta} = g_{\alpha\beta} - \frac{g_{0\alpha}g_{0\beta}}{g_{00}}
\end{eqnarray}
and also 
\begin{eqnarray} \label{Landau}
    g^{\mu\rho}h_{\rho\nu} = \delta^{\mu}_{\nu} \hspace{5mm} \implies [g]^{-1} = [h]
\end{eqnarray}
Thus, from (\ref{Omega}) we can write:
\begin{eqnarray} \label{Omega2}
    \text{Det}([\omega]) = \text{Det}([D])\frac{1}{\sqrt{\lambda_1\lambda_2\lambda_3}}
\end{eqnarray}
Now, 
\begin{eqnarray} 
        &[D]^T[g][D] = \text{Diag}(\lambda_1, \lambda_2, \lambda_3) \\
        &\hspace{-12mm}\implies \text{Det}([D]) = \frac{\sqrt{\lambda_1\lambda_2\lambda_3}}{\sqrt{\text{Det}([g])}} \label{Omega3}
\end{eqnarray}
Putting Det($[D]$) from (\ref{Omega3}) in (\ref{Omega2}) and finally using (\ref{Landau}), we have:
\begin{eqnarray} \label{Det}
    \text{Det}([\omega]) = \frac{1}{\sqrt{\text{Det}[g]}} = \sqrt{\text{Det}([h])}
\end{eqnarray}
Hence, combining (\ref{Det}) with (\ref{Master}), we can write:
\begin{eqnarray} \label{1P1}
    \Gamma(E) = \int_V d^3\mathbf{x}\sqrt{\text{Det}([h])}\bigg(\frac{4\pi}{3}\bigg)\bigg[\lambda + \sum_{\mu}\frac{\gamma_{\mu}^2}{4}\bigg]^{3/2}
\end{eqnarray}
where $V$ is the spatial extent of the box in which the particle resides. This is precisely the expression (\ref{1P}).

\section{$\Gamma(E)$ for Cuboidal Box} \label{AppB}
The expressions for $\Gamma(E)$ obtained in Sections-(\ref{Sec3}) and (\ref{Sec4}) were for a particle in a symmetrical spherical box. That led to curvature and acceleration corrections proportional to boundary area of the system. In this appendix, we show that such is not the general case by taking an asymmetric cuboidal box (side lengths along the Cartesian $x,y$ and $z$ axes are $L_x$, $L_y$ and $L_z$ respectively) and computing $\Gamma(E)$ for it. 
\newline \textbf{Note:} We will often refer to $x$ coordinate as $y^1$, $y$ as $y^2$ and $z$ as $y^3$ along with calling $L_x, L_y, L_z$ as $L_1, L_2, L_3$ respectively and they will be used interchangeably. 

\subsection{Single particle in an Accelerated box}\label{AppB1}
Consider the cuboidal box moving with a constant acceleration with its 3-acceleration directed along $x$-axis and magnitude given by $|\mathbf{a}|$
We have:
\begin{eqnarray}
    \Gamma(E) &=& \int_V d^3\mathbf{x}\bigg(\frac{4\pi}{3}\bigg)\bigg\{\frac{E^2}{(1+a_{\mu}x^{\mu})^2}-m
    ^2\bigg\}^{3/2} \nonumber \\
    &=& \bigg(\frac{4\pi}{3}\bigg) \int_{-L_y/2}^{L_y/2}dy \int_{-L_z/2}^{L_z/2}dz \int_{-L_x/2}^{L_x/2}dx \bigg\{\frac{E^2}{(1+|\mathbf{a}|x)^2}-m
    ^2\bigg\}^{3/2}  \label{B1}
\end{eqnarray}
Now, we consider the massless case ($m=0$) and hence (\ref{B1}) evaluates to:
\begin{eqnarray} \label{B2}
    \Gamma(E) = \bigg(\frac{4\pi}{3}\bigg)E^3V\bigg(1 - \frac{1}{4}|\mathbf{a}|^2L_x^2\bigg)^{-2}
\end{eqnarray}
Now, taking $|\mathbf{a}|L_x$ as a small parameter, equation (\ref{B2}) can be written as:
\begin{eqnarray} \label{B3}
    \Gamma(E) = \bigg(\frac{4\pi}{3}\bigg)E^3V\bigg(1 + \frac{1}{2}|\mathbf{a}|^2L_x^2 + \mathcal{O}(|\mathbf{a}|^4L_x^4)\bigg)
\end{eqnarray}
This equation (\ref{B3}) is the analogue of (\ref{Acc}) for a uniformaly accelerated cuboidal box. Note that only $L_x^2$ term appears and neither the total surface area, nor the cross-sectional area ($L_yL_z$) appears as the first order correction.  

\subsection{Single particle in a box in an arbitrary spacetime}\label{AppB2}
We have the expression (\ref{MasterEqn}) for $\Gamma(E)$ in FNC. We just need to compute it for the cuboidal box now. 
\newline Note that:
\begin{eqnarray}
    && \int_Vd^3\mathbf{y}\bigg(-\frac{1}{6}\sum_{\mu=1}^3R_{\mu\alpha\mu\beta}\bigg)y^{\alpha}y^{\beta} \nonumber\\&=& -\frac{1}{6}\int_{-L_x/2}^{L_x/2}dy^1\int_{-L_y/2}^{L_y/2}dy^2\int_{-L_z/2}^{L_z/2}dy^3R^{\mu}_{\alpha\mu\beta}y^{\alpha}y^{\beta}\nonumber \\
    &=& -\frac{1}{72}V\bigg(L_x^2\tensor{R}{^{\mu1}_{\mu1}}+L_y^2\tensor{R}{^{\mu2}_{\mu2}}+L_z^2\tensor{R}{^{\mu3}_{\mu3}}\bigg) \label{Note1}
\end{eqnarray}
Also similarly, 
\begin{eqnarray} \label{Note2}
    \int_Vd^3\mathbf{y}\tensor{R}{_{0\mu0\nu}}y^{\mu}y^{\nu} = \frac{1}{12}V\bigg(L_x^2\tensor{R}{^{01}_{01}}+L_y^2\tensor{R}{^{02}_{02}}+L_z^2\tensor{R}{^{03}_{03}}\bigg) 
\end{eqnarray}
and 
\begin{eqnarray} \label{Note3}
    \int_Vd^3\mathbf{y}a_{\mu}a_{\nu}y^{\mu}y^{\nu} = \frac{1}{12}V|\mathbf{a}|^2L_x^2
\end{eqnarray}
Hence, putting (\ref{Note1}), (\ref{Note2}) and (\ref{Note3}) in (\ref{Acc}), we arrive at:
\begin{eqnarray} \label{FinalExp}
    \Gamma(E) &=& \frac{4\pi}{3}(E^2-m^2)^{3/2}V\bigg[1-\frac{1}{72}\bigg(\sum_{\alpha=1}^3L_{\alpha}^2\tensor{R}{^{\mu\alpha}_{\mu\alpha}}\bigg) \nonumber \\  && + \bigg(\frac{E^2}{E^2-m^2}\bigg)\bigg\{\frac{3}{2}\bigg(\frac{1}{12}\sum_{\alpha=1}^3L_{\alpha}^2\tensor{R}{^{0\alpha}_{0\alpha}} + \frac{1}{4}|\mathbf{a}|^2L_x^2\bigg)\bigg\} \nonumber \\ &&+\frac{1}{8}\bigg(\frac{E^2}{E^2-m^2}\bigg)^2|\mathbf{a}|^2L_x^2\bigg]
\end{eqnarray}
This equation (\ref{FinalExp}) is the analogue of (\ref{Big}) for a single particle in a cuboidal box.
\newline One can also readily check that by putting $m=0$ and curvature terms to zero, (\ref{FinalExp}) reduces to (\ref{B3}).

\section{Approximate Timelike Killing field in FNC} \label{Appendix-C}
\subsection{Proof of Approximate KVF equations}\label{Appendix-C.1}
In this appendix, we aim to prove the equations (\ref{KVF1}) and (\ref{KVF2}). To do that we note that:
\begin{eqnarray}
    \nabla_i\xi_{j} + \nabla_j\xi_i = \mathcal{L}_{\xi}g_{ij} = g_{ia}\partial_j\xi^a + g_{ja}\partial_i\xi^a + \xi^a\partial_ag_{ij}
\end{eqnarray}
and as $\xi = (1,0,0,0)$, hence:
\begin{eqnarray}
    \nabla_i\xi_{j} + \nabla_j\xi_i = \partial_0g_{ij}
\end{eqnarray}
Now, from (\ref{FNC}), we note that:
\begin{eqnarray}
    \partial_0g_{00} = -2\dot{a}_{\mu}y^{\mu} - 2\dot{a}_{\mu}a_{\nu}y^{\mu}y^{\nu} + \dot{R}_{0\mu0\nu}y^{\mu}y^{\nu} +\mathcal{O}(y^3, \partial R) \label{1}\\
    \partial_0g_{0\mu} = -\frac{2}{3}\dot{R}_{0\rho\mu\nu}y^{\rho}y^{\nu} + \mathcal{O}(y^3, \partial R) \label{2}\\
    \partial_0g_{\mu\nu} = - \frac{1}{3}\dot{R}_{\mu\rho\nu\sigma}y^{\rho}y^{\sigma} + \mathcal{O}(y^3, \partial R) \label{3}
\end{eqnarray}
Hence for $\dot{a}^i = 0$ we recover (\ref{KVF1}) and for $\dot{a}^i \neq 0$, (\ref{KVF2}) holds.

\subsection{Approximate KVF: A Schwarzschild Example}\label{C.2}
From equations-(\ref{1} - \ref{3}), it is obvious that time derivative terms of Riemann tensor components along the timelike trajectory of the centroid of the box make the KVF approximate (i.e. not exact). Now suppose $x^{i}(\tau_0)$ is a point on the trajectory of the centroid of the box corresponding to proper time $\tau_0$. Let the corresponding point on the trajectory at some other point in proper time $\tau$ is $x^i(\tau)$ and we further denote $\tau - \tau_0 = \Delta\tau$. Now we want to put a constraint over this $\Delta\tau$ for which the approximation of KVF (and hence the approximation of conserved energy) is valid. We will denote a generic component of Riemann tensor as $\mathscr{R}$ instead of $R_{abcd}$. Now one has:
\begin{eqnarray}
    \mathscr{R}(x(\tau)) = \mathscr{R}(x(\tau_0)) + \dot{\mathscr{R}}(x(\tau_0))\Delta\tau + \mathcal{O}(\Delta\tau^2)
\end{eqnarray}
and hence we need the dimensionless quantity $\Big|\frac{\dot{\mathscr{R}}}{\mathscr{R}}\Delta\tau\Big| << 1$. Hence the timescale for which energy will be approximately conserved is
\begin{eqnarray} \label{Approx}
    \Delta\tau << \Big|\frac{\mathscr{R}}{\dot{\mathscr{R}}}\Big|
\end{eqnarray}
To make this condition (\ref{Approx}) a bit more concrete we give an example of a radial geodesic (i.e. centroid of the box freely falling radially) in Schwarzschild spacetime. FNC computations for such geodesics are done in (\cite{Misner}). The radial geodesic equations for Schwarzschild radial coordinate $r$ and proper time $\tau$ obtained in (\cite{Misner}) are:
\begin{eqnarray}
    r &=& \frac{1}{2}r_0(1+\cos{\omega}) \\
    \tau &=& \frac{1}{2}r_0\sqrt{\frac{r_0}{2M}}(\omega + \sin{\omega})
\end{eqnarray}
in terms of a cycloid paramter $\omega \in (0,\pi)$, the Schwarzschild mass parameter $M$ and initial radial point $r_0$. Now apart from numeric constants, the non-zero Riemann curvature terms behave as $R \sim \frac{M}{r^3}$. So, we have
\begin{eqnarray}
    \Big|\frac{\mathscr{R}}{\dot{\mathscr{R}}}\Big| \sim \Big|\frac{r}{\dot{r}}\Big| \sim \frac{r_0^{3/2}}{M^{1/2}} \sim \frac{1}{\sqrt{\mathscr{R}_0}}
\end{eqnarray}
where $\mathscr{R}_0$ is Riemann curvature tensor at the initial point of the particle. Hence for this case, the time-scales for which the KVF approximation (and hence the notion of conserved energy) is valid is given by:
\begin{equation}
    \Delta\tau << \frac{1}{\sqrt{\mathscr{R}_0}} \sim L_{\mathscr{R}}
\end{equation}
where $L_{\mathscr{R}}$ is the curvature length-scale at the initial point of the particle.

\section*{Competing Interests}
The authors declare that they have no competing interests.

\section*{Funding}
This research received no external funding.

\printbibliography

\end{document}